\documentclass{aa}  

\usepackage{graphicx}
\usepackage[varg]{txfonts}
\usepackage{natbib} 
\bibpunct{(}{)}{;}{a}{}{,} 
\usepackage{longtable}   
\usepackage{xcolor}
\begin{document}

   \title{H$\alpha$ emission-line stars in molecular clouds}
  
   \subtitle{III. Canis Major\thanks{Table 2 is only available in electronic form
at the CDS via anonymous ftp to cdsarc.u-strasbg.fr (130.79.128.5)
or via http://cdsweb.u-strasbg.fr/cgi-bin/qcat?J/A+A/}}

   \author{Bertil Pettersson
          \inst{1}
          \and
          Bo Reipurth\inst{2}
          }

   \institute{Observational Astronomy, Division of Astronomy and Space Physics, Department of Physics and Astronomy, Uppsala University, Box 516,
           SE-751 20 Uppsala, Sweden\\
              \email{bertil.pettersson@physics.uu.se}
         \and
             Institute for Astronomy, University of Hawaii at Manoa, 
          640 North Aohoku Place, Hilo, HI 96720, USA\\
             \email{reipurth@ifa.hawaii.edu}
             }

   \date{Received July 16 2017; Accepted July 24 2019}

\abstract{

A deep objective-prism survey for H$\alpha$ emission stars towards
  the Canis Major star-forming clouds was performed. A total of
  398 H$\alpha$ emitters were detected, 353 of which are new
  detections. There is a strong concentration of these H$\alpha$
  emitters towards the molecular clouds surrounding the CMa~OB1
  association, and it is likely that these stars are young stellar
  objects recently born in the clouds.  An additional population of
  H$\alpha$ emitters is scattered all across the region, and probably
  includes unrelated foreground dMe stars and background Be
  stars. About 90\% of the H$\alpha$ emitters are detected by WISE, of which 75
  \% was detected with usable photometry. When plotted in a WISE colour-colour
  diagram it appears that the majority are Class~II YSOs. Coordinates
  and finding charts are provided for all the new stars,  and
  coordinates for all the detections. We searched the Gaia-DR2 catalogue and
  from 334 H$\alpha$ emission stars with useful parallaxes, we selected a subset of 98
  stars that have parallax errors of less than 20\% and nominal
  distances in the interval 1050 to 1350~pc
that surrounds a strong peak at 1185~pc in the distance
distribution. Similarly, Gaia distances were obtained for 51 OB-stars
located towards Canis Major and selected with the same parallax errors as the
H$\alpha$ stars. We find a median distance for the OB stars of
1182~pc, in excellent correspondence with the distance from the
H$\alpha$ stars. Two known runaway stars are confirmed as members of the association. Finally, two new Herbig-Haro objects are identified.
  }


  \keywords{
stars: formation -- stars: pre-main sequence -- stars: emission-line, Be
}

\maketitle

\defcitealias{ReipPett04}{Paper~I}
\defcitealias{PettArmo14}{Paper~II}

\section{Introduction}

Objective prism surveys for H$\alpha$ emission stars
were the first technique to identify large populations of young stars
\citep[e.g.][]{Haro53,Herbig54}. Over time, many of the closer star
forming regions were surveyed \citep[e.g.][]{Schwartz77}. These early
surveys were all done using photographic plates, but with the advent of
CCDs combined with grisms, much deeper surveys were possible, albeit over much
smaller areas \citep[e.g.][]{Herbig98, HerbDahm06}. We performed 
a major survey of H$\alpha$ emission stars in southern
star-forming regions based on a large set of wide-field photographic
films obtained in the late 1990s at the European Southern Observatory (ESO) Schmidt telescope with
objective prism before it was closed down. In Paper~I \citep{ReipPett04} 
we studied NGC~2264 and presented 357 H$\alpha$ emission-line
stars, of which 244 were new. In Paper~II \citep{PettArmo14} we
studied the M42 region and presented 1699 H$\alpha$ emitters, of which
1025 were new detections. In this paper we present the results of our
survey of the Canis Major star-forming region.

   \begin{figure*}
   \centering
   \includegraphics[width=170mm]{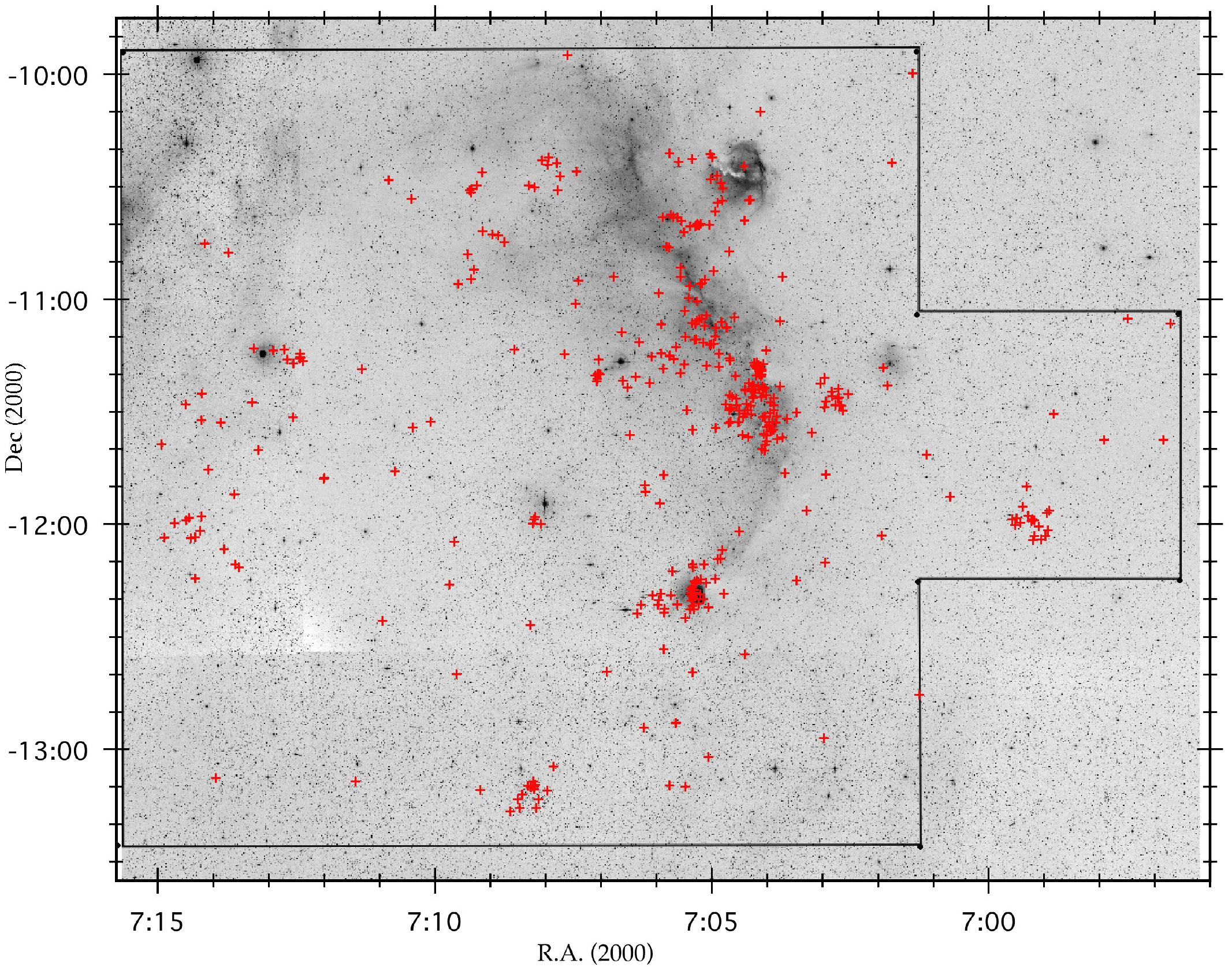}
      \caption{Red image from DSS2 showing surveyed area bordered by a black line. The distribution of all detected emission-line stars is shown, each marked as a red cross
              }
         \label{fig1}
   \end{figure*}

CMa OB1 was first recognised as a stellar association by \citet{Ambart47},
 and more members were identified by \citet{Clar74x,Clar74y}. Within
the boundaries of CMa~OB1, \citet{vdBergh66} found a number of stars
associated with reflection nebulae, a grouping that he called CMa~R1.
These stars were studied in more detail by \citet{Racine68}. The
relation between CMa~OB1 and CMa~R1 was disputed for some time, but
it is now recognised that they are part of the same association of
young stars. The distance to CMa~OB1 has been estimated multiple
times, with results varying from 690~pc 1150~pc.  The large spread in
values is not surprising given that towards Canis Major we are looking
at a shallow angle down the Orion-Cygnus arm for almost 2~kpc, with
the Perseus arm further behind. In the following we
present an accurate Gaia-DR2 determination of the distance to the
young stars in this region. 

The HII region excited by the OB stars, Sh2-296, is particularly
pronounced towards the west, where it is bounded by a ridge of
molecular clouds that are outlined in the CO studies of, for example 
\citet{DameHart01}, and in the extinction studies of \citet{Doba11} and
\citet{LombAlve11}.  \citet{HerbAsso77} suggested that star
formation in the ring-like structure of interstellar matter
surrounding CMa~OB1 and CMa~R1 has been triggered by a supernova. This
was supported by the study of \citet{ComeTorr98}, who used
Hipparcos to investigate the motions of the O and B stars, and found a
new runaway star, HD~57682, that $\sim$10$^6$~yr ago passed through
the estimated centre of expansion of the association. \citet{ShevEzhk99} 
identified many probable members of the association.

Young low-mass stars have formed in the Canis Major clouds. The most
famous is the FUor Z~CMa \citep[e.g.][]{ HartKeny89}, which is
located in the large cloud complex, L1657, to the west of the OB
association.  A number of H$\alpha$ emission-line stars have been
found in Canis Major. \citet{Wack70} discovered some in his early-type,
emission-line survey across the sky, and \citet{StepSand77} included
Canis Major in their general H$\alpha$ emission survey along the Milky
Way. The first H$\alpha$ emission star survey dedicated to the Canis
Major region was by \citet{WiraKogu86}, who catalogued 129
emission-line stars. Later, \citet{SugiTamu95} observed a small
bright-rimmed core, BRC~27, along the L1657 cloud edge and found
infrared (IR)-excess stars there, and subsequently \citet{OgurSugi02}
 found 32 H$\alpha$ emission stars towards the same cloud core.
\citet{SoarBica02,SoarBica03} used the 2MASS catalogue to study four
small clusterings of young stars, including BRC~27.
\citet{GregMont09} and \citet{FernGreg15} used X-ray data from ROSAT
to identify young stars across a $\sim$5~sq.  degree field, mainly
concentrated in two groupings, one near Z~CMa and another near
GU~CMa. \citet{MallOjha12} presented a multi-wavelength study of the
small HII region Sh2-297 at the southern tip of the L1657 cloud, and
showed that multiple populations of young stars with ages from 1 to
20~Myr co-exist in this region. \citet{RebuJohn13} used Spitzer data
to study the young stars in BRC 27. \citet{EliaMoli13} used Herschel
data to study star formation in a large area along the Galactic plane
including Canis Major.  Most recently, \citet{FiscPadg16} used the
Wide-Field Infrared Survey Explorer (WISE) to identify many young
stars across a very wide area in Canis Major.  Below we compare our
results to these various studies. 
Reviews of the literature on the Canis Major star-forming region have
been presented by \citet{Herbig91} and \citet{Greg08}.

\section{Observations}

We have used the ESO 1m Schmidt telescope at the La Silla Observatory
to obtain objective prism exposures of the Canis Major star-forming
region.  The objective prism gives a dispersion of 800 \mbox{\AA\
  mm$^{-1}$} at H$\alpha$, as previously described in more detail in
\citetalias{ReipPett04}.  Sensitised Kodak 4415 emulsion on spectral films was used.
A series of exposures were obtained from November 1996 to late January
1997, and from these the seven best films were selected, all taken
during photometric nights with good seeing. Exposure times were chosen
to vary between 10 min and 150 min in order to detect H$\alpha$
emission in faint as well as bright stars.  A RG~630 filter was used
which provides a spectral range from 630 to 690 nm, a narrow range
centred on the H$\alpha$ line.  The large films of 30 cm $\times$ 30
cm cover a field of 5\fdg4 $ \times$ 5\fdg4 on the sky. All exposures
were centred on the position
$\alpha$:~7$^\mathrm{h}$04$^\mathrm{m}$,
$\delta$:~-11$\degr$ 28\arcmin\ (J2000).  
Table~ \ref{tbl1} lists the films employed in the present study.  The Schmidt
films were visually inspected with the use of a binocular microscope
in search of emission against the continuum of the stars.

\begin{table}
\caption{List of Schmidt films }             
\label{tbl1}      
\centering                          
\begin{tabular}{c c c c c}        
\hline\hline                 
Date &  No. & Exp. time & Filter & Seeing \\ 
     &      & (minutes) &        & (\arcsec) \\    
\hline                        
1996 Nov 14 & 12841B & 40 & RG630 & 0.85  \\
1996 Dec 13 & 12897B &130 & RG630 & 0.8  \\
1996 Dec 14 & 12900B & 10 & RG630 & 0.78  \\
1996 Dec 31 & 12924B &150 & RG630 & 1.0  \\
1997 Jan 08 & 12944B & 40 & RG630 & 0.8  \\
1997 Jan 09 & 12947B & 90 & RG630 & 0.8  \\
1997 Jan 29 & 12969B &  90 & RG630 & 0.8  \\
\hline                                   
\end{tabular}
\end{table}

The 2nd Digitised Sky Survey (DSS2),
which provides B,R,I images with a plate scale of 1$''$ pixel$^{-1}$,
was used to obtain initial coordinates for the stars. The images were
retrieved in FITS format and had astrometric information in their
headers. The task SAOimage was used to determine the
coordinates of each star visually identified in the finding charts.
R-images were used to measure the coordinates for the stars.

These initial coordinates were searched (within a radius of 2$''$)
among the stars of the 2 Micron All Sky Survey (2MASS) catalogue in
order to obtain near infrared J H K$_s$ magnitudes.  The 2MASS
coordinates have expected uncertainties of 0.1$''$, better than what we
achieved with the DSS2 images. Therefore, the present survey uses the
2MASS coordinates.

\section{Results}

\subsection{Survey}

We have detected 398 H$\alpha$ emission stars within
an area of approximately 14 square-degrees as outlined in
Fig.~\ref{fig1}.  Among these, 45 H$\alpha$
emitters were previously known, so 353 objects are
new. Finding charts for the new objects are presented in Appendix A. Originally our survey was focused on a rectangular area slightly
larger than 3$\times$3 degrees centred on the CMa OB association, but
a concentration of H$\alpha$ emitters was spotted to the west, and so
the survey was extended to that region, see below for further details.

\begin{table*}
\caption{ H$\alpha$ emission-line stars in CMa OB 1}
\label{tbl2}
\centering
{
\tabcolsep 3pt
\resizebox{182mm} {!} {
\begin{tabular}{llcccccccrrrccl}
\hline\hline
ESO-Ha\tablefootmark{a} & Other\tablefootmark{b} &
RA(2000)\tablefootmark{c} & Dec(2000)$^c$ & H$\alpha$\tablefootmark{d} &  
$G$\tablefootmark{e} & $J^\mathrm{f}$ & $H^\mathrm{f}$ & $K_{s}$\tablefootmark{f} & [3.4]\tablefootmark{g} &  [4.6]\tablefootmark{g} & [12]\tablefootmark{g}& dist \tablefootmark{h}(pc)& dist errors\tablefootmark{h} &Notes
\tablefootmark{i} \\
\hline
1979 &  & 7:04:52.38 & $-$11:17:58.5 & 5 & 18.79 & 15.57 & 14.51 & 13.63 & 12.26 & 11.57 & ~9.38 & 1238& +1210,-409&   \\
1980 &  & 7:04:53.66 & $-$10:34:13.8 & 5 & 18.39 & 13.29 & 11.55 & 10.07 & ~8.36 & ~7.41 & ~5.26 & 1420 &+955,-407&  NW component of bin \\
1981 &  & 7:04:53.80 & $-$12:09:19.2 & 4 & 18.02 & 14.57 & 13.58 & 13.17 & 12.24 & 11.75 & ~8.41 & 1259& +391,-241&   \\
1982 &  & 7:04:54.10 & $-$10:27:01.9 & 4 & 18.13 & 15.03 & 14.30 & 13.98 & 13.38 & 12.88 & 11.46: & 1347& +386,-245&   \\
 & Kiso31 & 7:04:55.78 & $-$11:34:17.6 & 3 & 13.60 & 11.53 & 10.87 & 10.48 & ~9.95 & ~9.61 & ~8.23 & &  \\
1983 &  & 7:04:55.86 & $-$11:07:38.7 & 5 & 18.34 & 14.24 & 13.28 & 12.97 & 12.01 & 11.68 & ~9.08 & ~548 &+99,-73&  W comp of close pair \\
1984 &  & 7:04:56.24 & $-$10:36:34.5 & 5 & 19.61 & 15.37 & 14.20 & 13.35 & 12.49 & 11.62 & 10.68 & ~742&+498,-213&  No continuum \\
1985 &  & 7:04:56.53 & $-$11:09:52.3 & 5 & 18.42 & 15.47 & 14.85 & 14.36 &  &  & & 1685 &+1234,-501&   \\
1986 &  & 7:04:56.70 & $-$12:14:32.9 & 4 & 17.26 & 14.06 & 12.94 & 12.34 & 11.47 & 11.00 & ~8.48 & 1016 &+131,-104&   \\
1987 &  & 7:04:57.92 & $-$10:52:29.5 & 2 & 17.50 & 13.28 & 12.24 & 11.79 & 11.18 & 10.77 & ~8.66 & & &  \\
1988 &  & 7:04:58.05 & $-$11:11:48.9 & 1 & 15.99 & 13.49 & 12.76 & 12.60 & 12.41 & 12.32 & 10.53 & 1185 &+96,-82&   \\
1989 &  & 7:04:59.79 & $-$10:22:13.4 & 2 & 18.15 & 15.03 & 14.08 & 13.81 & 13.47 & 13.63 & ~9.21 & 1516 &+579,-328&   \\
1990 &  & 7:05:00.27 & $-$11:12:01.0 & 3 & 17.18 & 14.32 & 13.46 & 13.21 & 12.99 & 12.74 & 11.06: & 1073 &+177,-133&   \\
1991 &  & 7:05:01.37 & $-$10:28:00.8 & 3 & 17.39 & 14.63 & 13.78 & 13.48 &  &  & & ~953 &+110,-89&  Close to brighter star \\
1992 &  & 7:05:01.53 & $-$10:21:18.3 & 4 & 17.97 & 14.70 & 13.74 & 13.42 & 12.97 & 12.41 & ~9.94 & ~914 &+155,-116&   \\

\hline
\end{tabular}
}
\tablefoot{
The full table with all the 398 H$\alpha$ emission-line stars is available 
electronically at CDS. A few lines are reproduced here only for guidance regarding 
format and content.
\tablefoottext{a}{The numbers fall into two groups (1835-2123 and 2706-2769).}
\tablefoottext{b}{Kiso: \citet{WiraKogu86}; SS: \citet{StepSand77};  LkH$\alpha$: \citet{Herbig60}; OSP: \citet{OgurSugi02}.}
\tablefoottext{c}{Positions extracted from the 2MASS All-Sky Catalogue.}
\tablefoottext{d}{The H$\alpha$ strength is defined so 1 is weak emission against 
a strong continuum and 5 is strong emission against a weak or invisible continuum. Hyphenated 
values may represent either variability and/or uncertainty in the estimate.}
\tablefoottext{e}{Broad band magnitudes from Gaia.}
\tablefoottext{f}{$JHK_s$ magnitudes extracted from the 2MASS All-Sky Catalogue.}
\tablefoottext{g}{Magnitudes from  AllWISE.}
\tablefoottext{h}{Distances and errors extracted from Gaia-DR2}
\tablefoottext{i}{Stars with uncertain photometry in one or several bands are flagged with a ':'. Those stars are not considered in discussions or plots regarding photometry.}\\
} 
} 
\end{table*}

The 398 H$\alpha$ emission stars detected are listed in Table~2,
available at CDS. The table gives the following columns: (1) The
ESO-H$\alpha$ number, extending the numbering of our previous papers
in this series. It should be noted that we do not give ESO-H$\alpha$ numbers to
already known H$\alpha$ stars in order to limit the proliferation of
names for the same object. The numbers are in two groups,
ESO-H$\alpha$ 1835 to
  2123, and ESO-H$\alpha$ 2706 to 2769, because we
initially had planned another order for this series of survey papers.
(2) Previously used designations for stars that have already been
discovered in earlier surveys. (3-4) The 2000-coordinates as extracted
from the 2MASS catalogue. (5) The H$\alpha$ strength on a scale from 1
to 5, where 1 is weak emission relative to a stronger continuum and 5
is strong emission against a weaker (or in some cases invisible)
continuum.  Hyphenated values may represent either variability and/or
uncertainty in the estimate.  (6) Optical broad band G
  magnitudes extracted from the Gaia-DR2 catalogue.
  (7-9) JHK$_{s}$ magnitudes extracted from the 2MASS All-Sky Catalogue.
  (10-12) [3.4], [4.6] and [12] micron magnitudes from AllWISE. (13) Distance from Gaia-DR2. (14) Distance errors from Gaia-DR2. (15) Notes to
  individual stars.

\begin{figure}
   \centering
   \includegraphics[width=\columnwidth]{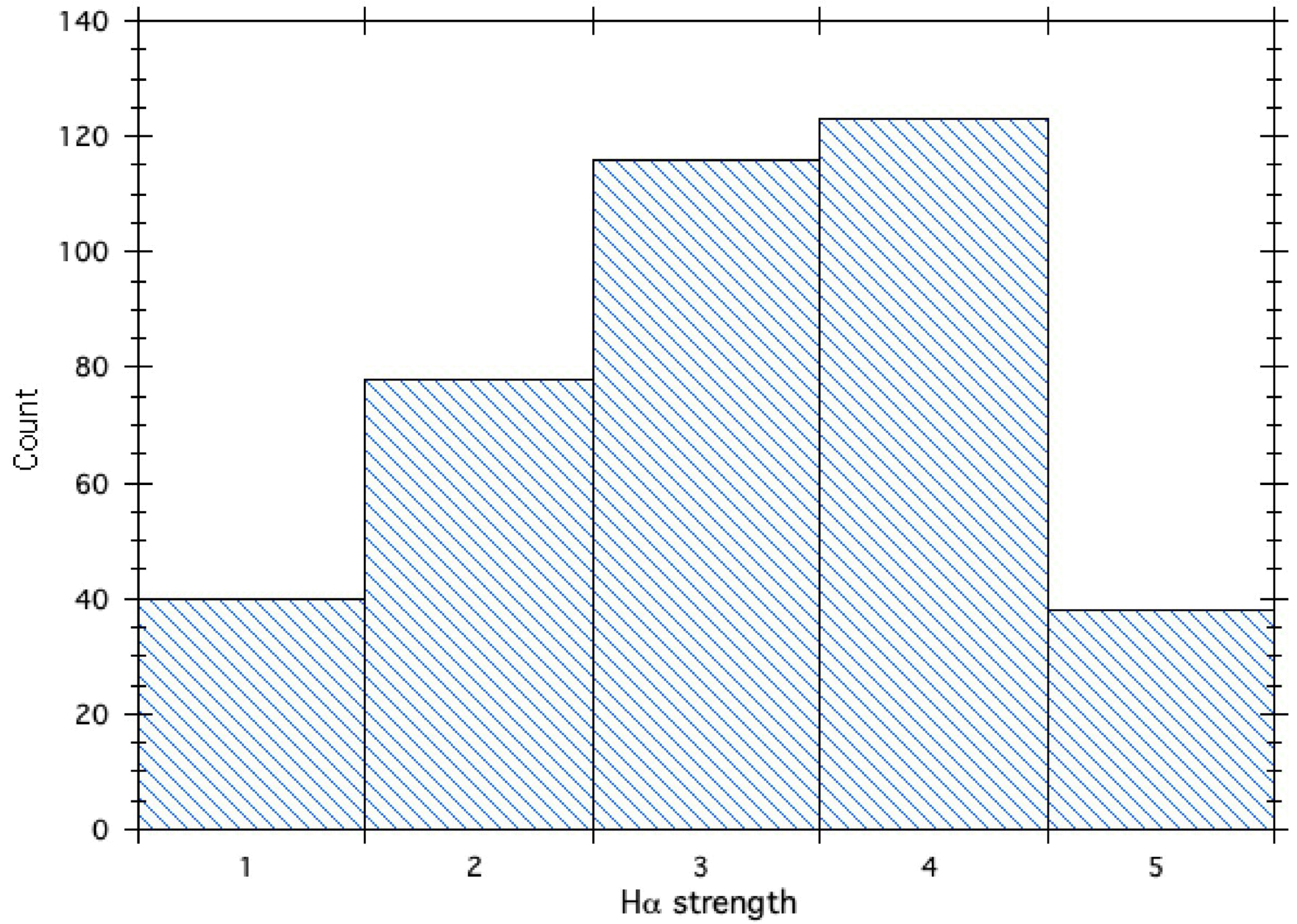}
      \caption{Distribution of emission-line strengths with 1 indicating weak emission relative to the continuum and 5 very strong.
              }
         \label{fig2}
   \end{figure}

Figure~\ref{fig2} shows the numbers of H$\alpha$ emitters distributed
across the five relative strengths of H$\alpha$ emission. 60\% of the stars are
found in strength 2 and 3. Strength 1 may be incomplete, since marginal
detections have been removed to ensure, as much as possible, that the
list contains only bona fide H$\alpha$ emitters, even at the expense
of completeness. Seven films were examined, and in some cases the
H$\alpha$ strength was found to vary, and in such cases the lower
value is shown in Fig.~\ref{fig2}. 

\begin{figure}
   \centering
   \includegraphics[width=\columnwidth]{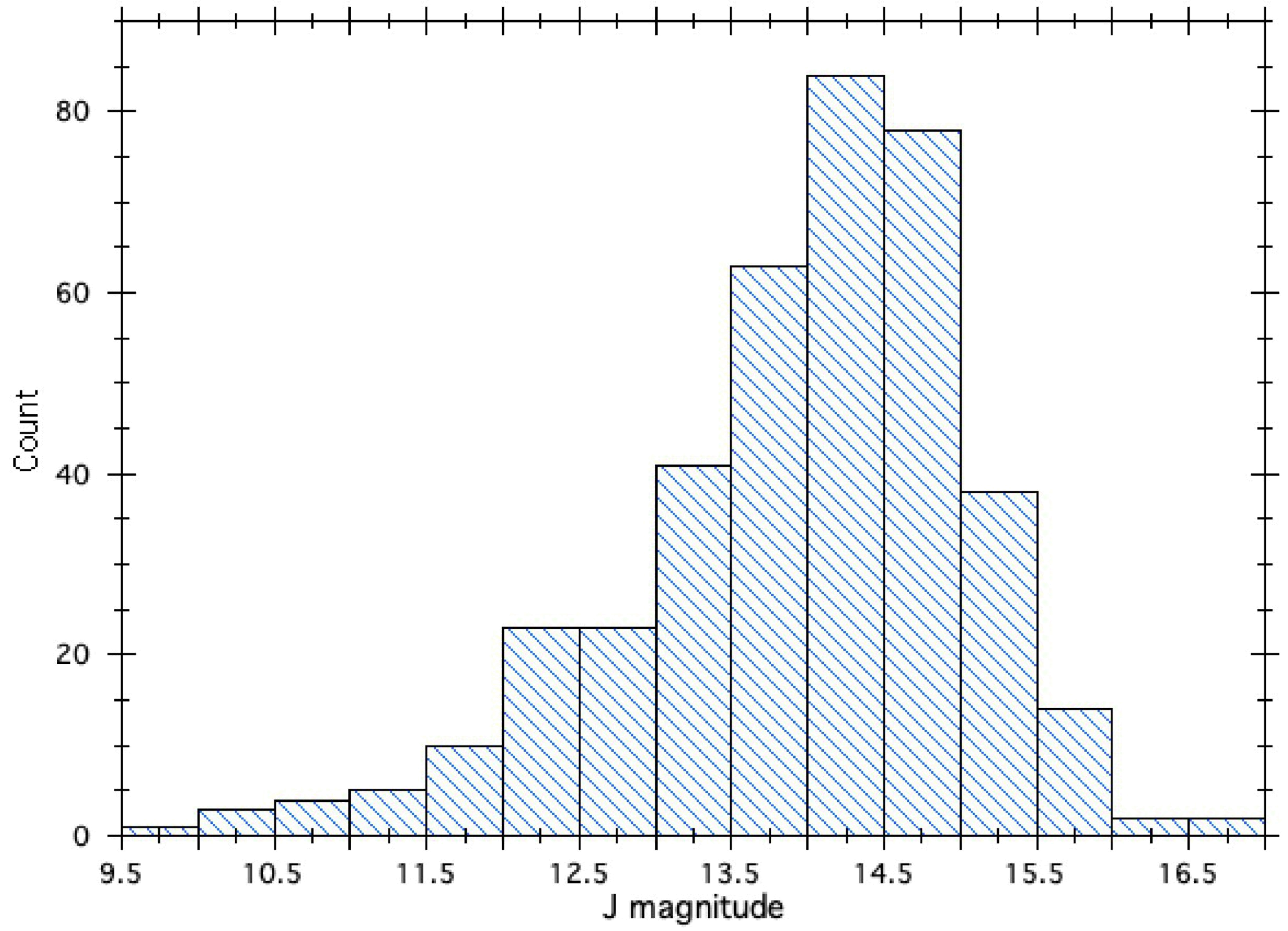}
      \caption{Distribution of J magnitudes of all detected stars as extracted from the 2MASS catalogue.
              }
         \label{fig3}
   \end{figure}

The brightness distribution of the H$\alpha$ emitters is shown in
Fig.~\ref{fig3}, where the 2MASS J-magnitude is used. The steep
drop at magnitudes fainter than 15 is likely to imply that our survey
is complete only to about that magnitude. Four stars were too faint to
be detected in 2MASS~J.

%


\subsection{Distribution} 

\begin{figure*}
   \centering
   \includegraphics[width=\textwidth]{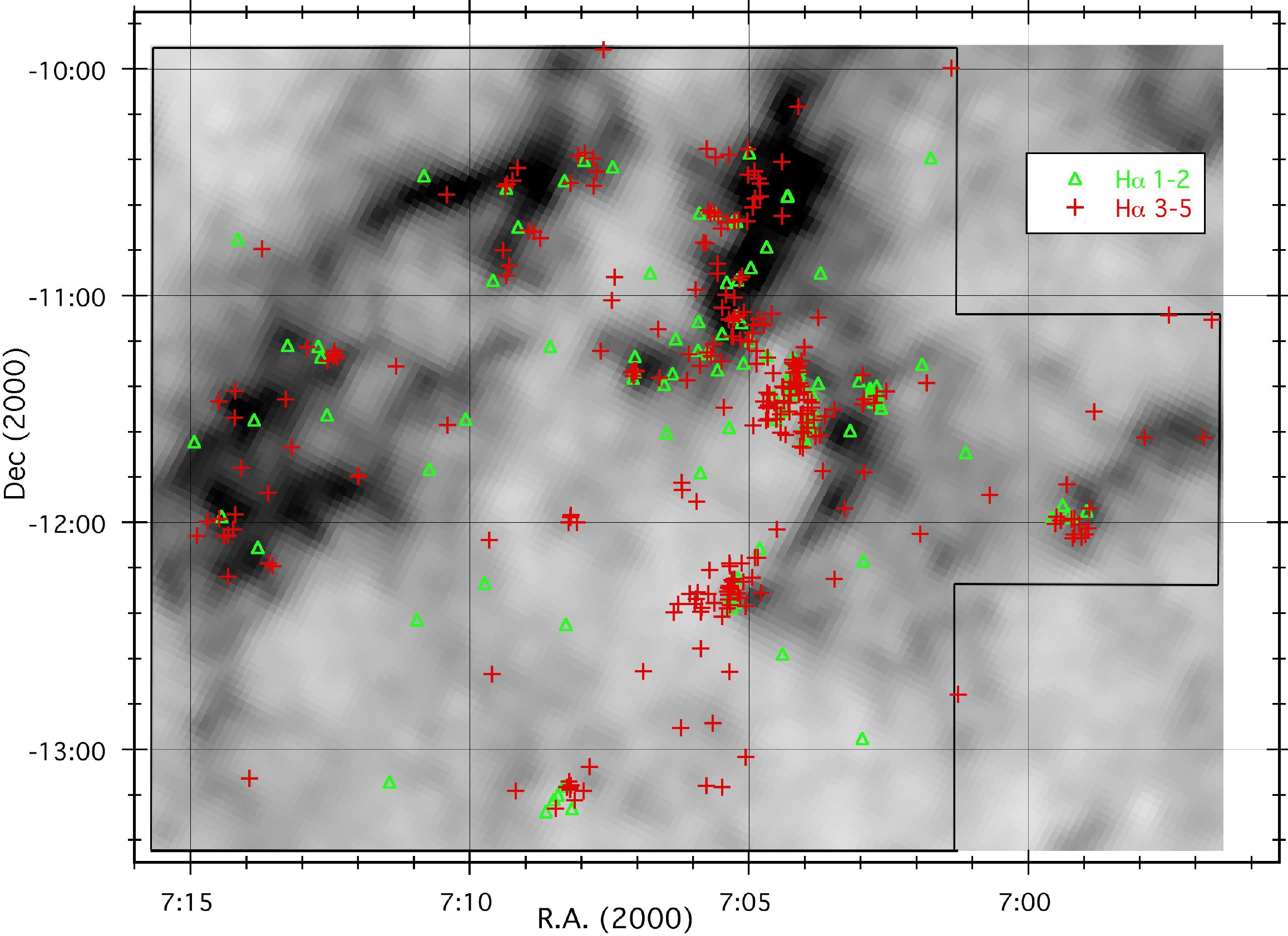}
      \caption{Distribution of detected emission-line stars superposed on an Av extinction map from \citet{DobaUeha05}.
                   Stars with H$\alpha$ strength 1-2 are represented by green triangles and the stronger emitters with H$\alpha$ strength 3-5 by red crosses.
              }
         \label{fig4}
   \end{figure*}

The H$\alpha$ emitting stars are not uniformly spread across the CMa
region, but their distribution shows significant substructure, with
stars concentrated in several groups. There is a clear association
between H$\alpha$ emitters and the eastern ridge of the bright-rimmed
clouds. The clouds surrounding the CMa OB1 association are difficult
to discern on the optical image in Fig.\ref{fig1} so we have used
the extinction measurements from \citet{DobaUeha05} to make an 
extinction map of the region, seen in Fig.~\ref{fig4}.

\begin{figure}
   \centering
   \includegraphics[width=\columnwidth]{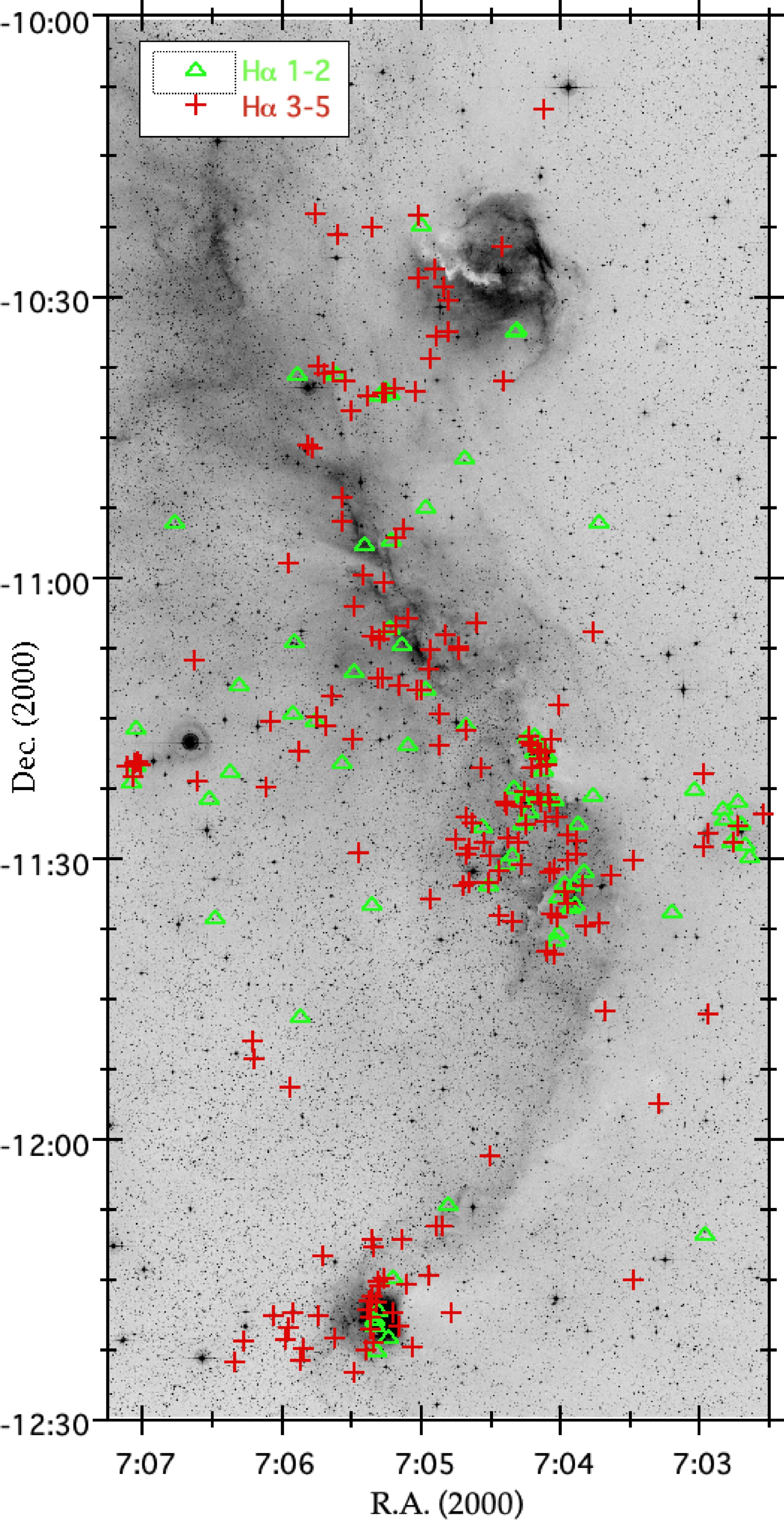}
      \caption{Enlargement of the distribution of H$\alpha$ emitters towards L1657.
      Stars with H$\alpha$ strength 1-2 are represented by green triangles and the stronger emitters with H$\alpha$ strength 3-5 by red crosses.
}
         \label{fig5}
   \end{figure}

The shell-like structure surrounding the main
group of OB stars is clearly outlined in extinction, the western side
of the shell forms the L1657 cloud, while the north-eastern side is
known as L1658 \citep{Lynds62}.  The large majority of H$\alpha$ emitters
are closely associated with these two cloud complexes, especially the
L1657 cloud. Figure~\ref{fig5} shows the distribution of H$\alpha$
emitters on an enlargement of the optical image of the L1657 cloud.
Further to the west, in an extension of the survey area, is another
cloud, known as TGU-1571 \citep{DobaUeha05}, which is also
associated with H$\alpha$ emitters. In addition to the clusterings of
stars associated with these three cloud complexes there are stars that
appear in regions of very low extinction, which may have three
%
   \begin{figure}
   \centering
   \includegraphics[width=\columnwidth]{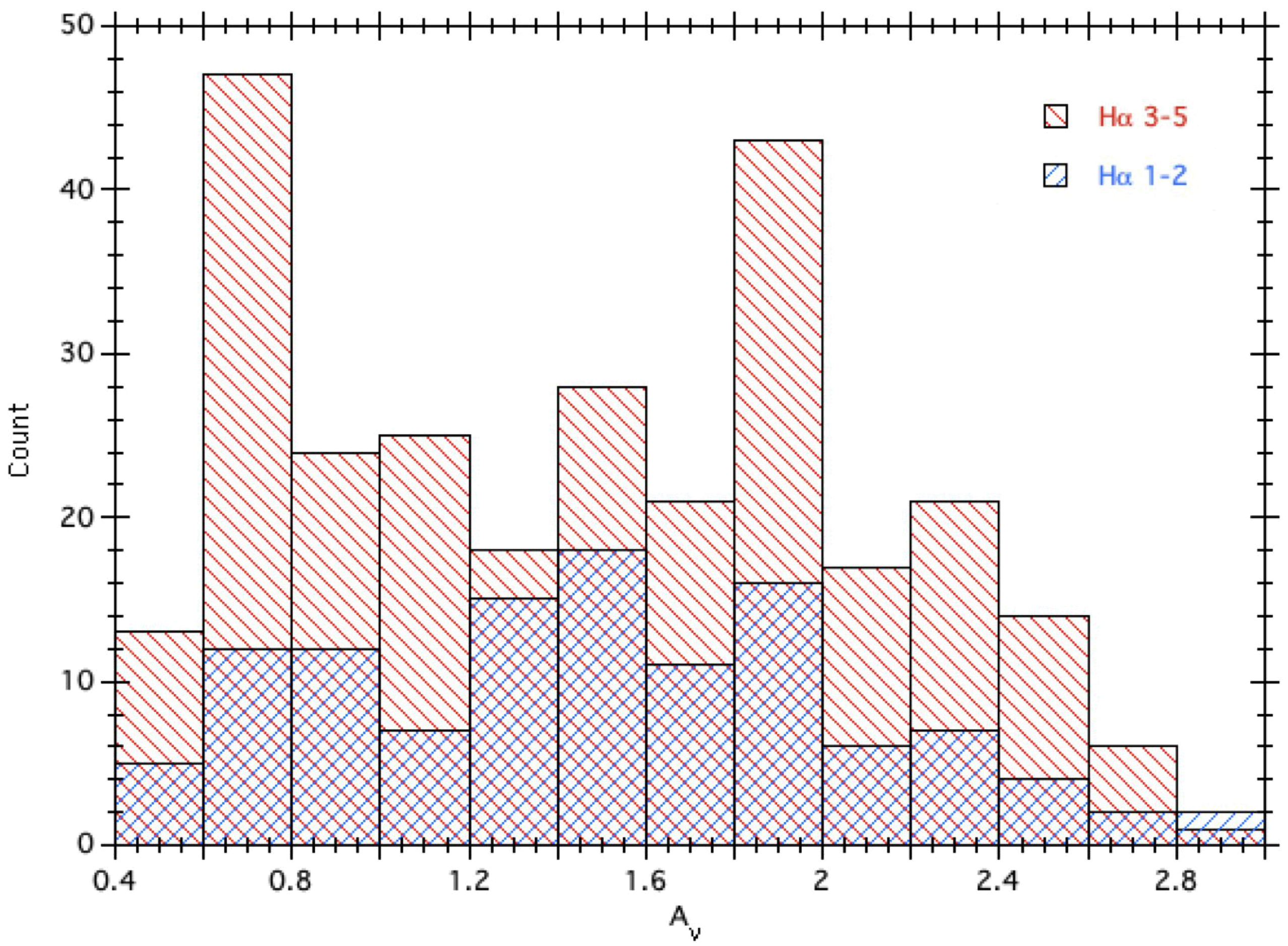}
      \caption{Relative distribution of H$\alpha$-emitters as function of the extinction
A$_V$ along the line of sight. The stronger-line stars do not show a
higher affinity of association with molecular clouds than the weaker
emission-line stars.
              }
         \label{fig6}
   \end{figure}
reasons. First, the stars could be born near the region where they are
found out of small clouds that are not easily detected with the
extinction mapping technique or, second, they could be slightly older stars
that have had time to wander away from their birth sites or, third,
they may represent a foreground or background population of unrelated
stars, such as dMe or Be stars, respectively. To investigate the
second possibility, we assume that a first generation of young stars
by now would have weaker H$\alpha$ emission and thus should be of
strength 1 and 2. In Fig.~\ref{fig6} we have plotted the H$\alpha$
emitters as function of the extinction A$_V$ in the general
neighbourhood of each star. The columns are divided into strength 1+2 (blue)
vs strength 3+4+5 (red). No obvious correlation between H$\alpha$
strength and local extinction is seen. Thus it appears that the scattered population of
H$\alpha$ emitters form a mix of isolated star formation and unrelated
emission-line stars of other types.

   \begin{figure}
   \centering
   \includegraphics[width=\columnwidth]{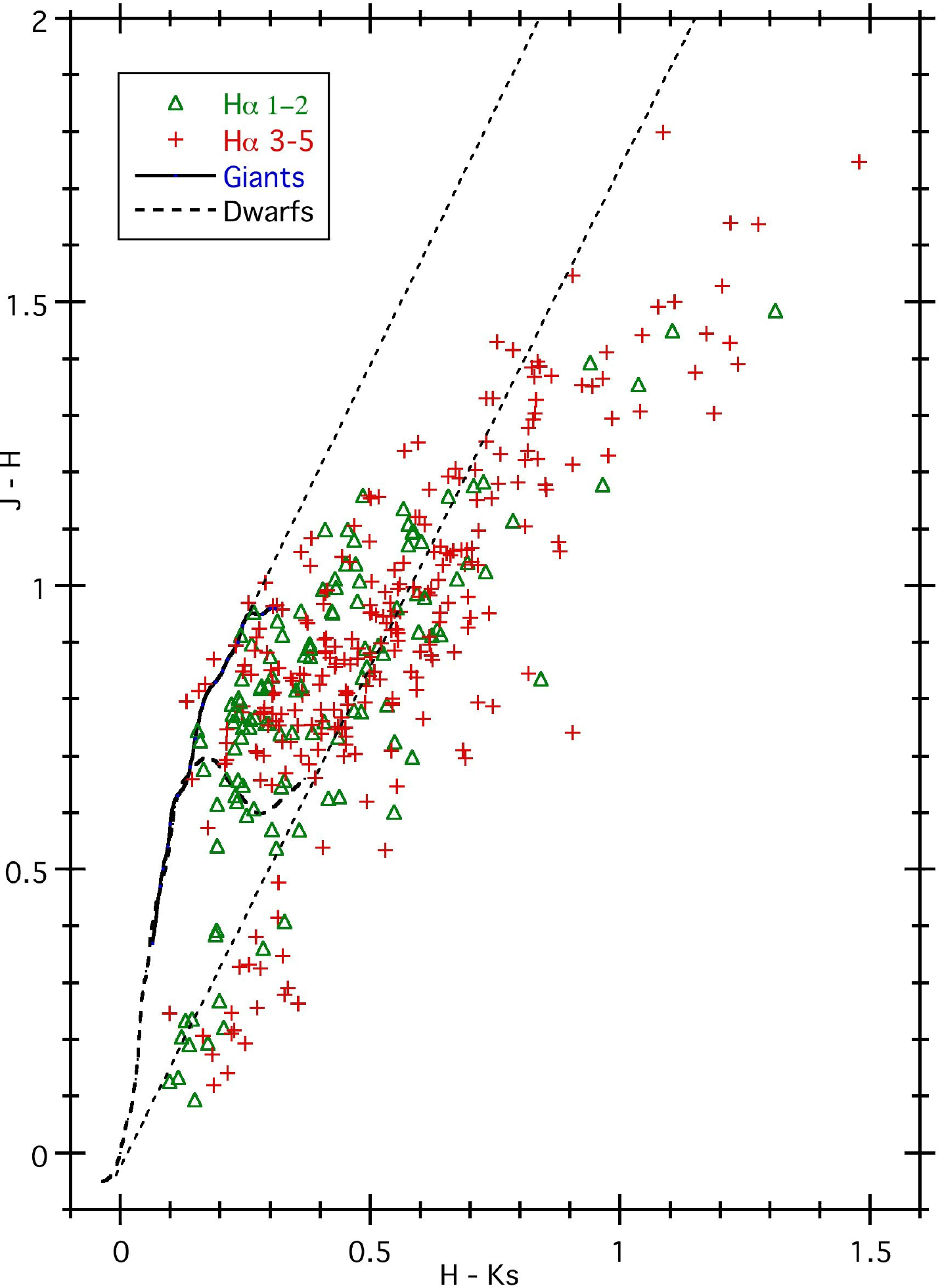}
      \caption{Near-IR colour-colour diagram based on 2MASS data showing all 
H$\alpha$ emitters in the survey with 2MASS detections in all three bands ($JHK_s$) (except upper limits).
Main sequence and giant loci from \citet{BB88}, corrected to the
2MASS photometric system \citep{CH01}, and the extinction law
from \citet{RL85} are used in the figure. 
              }
         \label{fig7}
   \end{figure}

\subsection{IR-excess stars}

We have studied the infrared properties of the H$\alpha$ emitters
using the 2MASS and WISE catalogues. Figure~\ref{fig7} shows a J-H
vs  H-K$_{s}$ colour-colour diagram of the H$\alpha$ emitters. The area
between the dashed lines indicate as usual the region where reddened
main sequence and giant stars can be found, and for which an infrared
excess cannot be unambiguously determined from JHK$_{s}$ photometry. Stars
to the right of this area cannot be just reddened main sequence or
giant stars, and they clearly have an infrared excess. About
44\% of the H$\alpha$ emitters have such
near-infrared excesses.  This compares to the 16\% of H$\alpha$
emitters with infrared excess in our identical survey of the general
neighbourhood of the Orion Nebula Cluster \citepalias{PettArmo14}.
This might indicate that the stars near the ONC have their
circumstellar material more UV-eroded than in the CMa~OB1 region.  We
find that the ratio of strong-vs-weak emission-line stars (strength
3-4-5 vs strength 1-2) for the total population of
398 H$\alpha$ emitters is 2.3, while
the same ratio for only the infrared excess stars is 3.4. 
As expected, the infrared excess stars are more likely to have
stronger H$\alpha$ emission, but at the same time there is clearly not
a linear relation between the two quantities. This is not unexpected,
since the infrared excess is a measure of the reservoir of
circumstellar material near the star whereas the H$\alpha$ emission is
an indicator of current accretion activity.
   \begin{figure}
   \centering
   \includegraphics[width=\columnwidth]{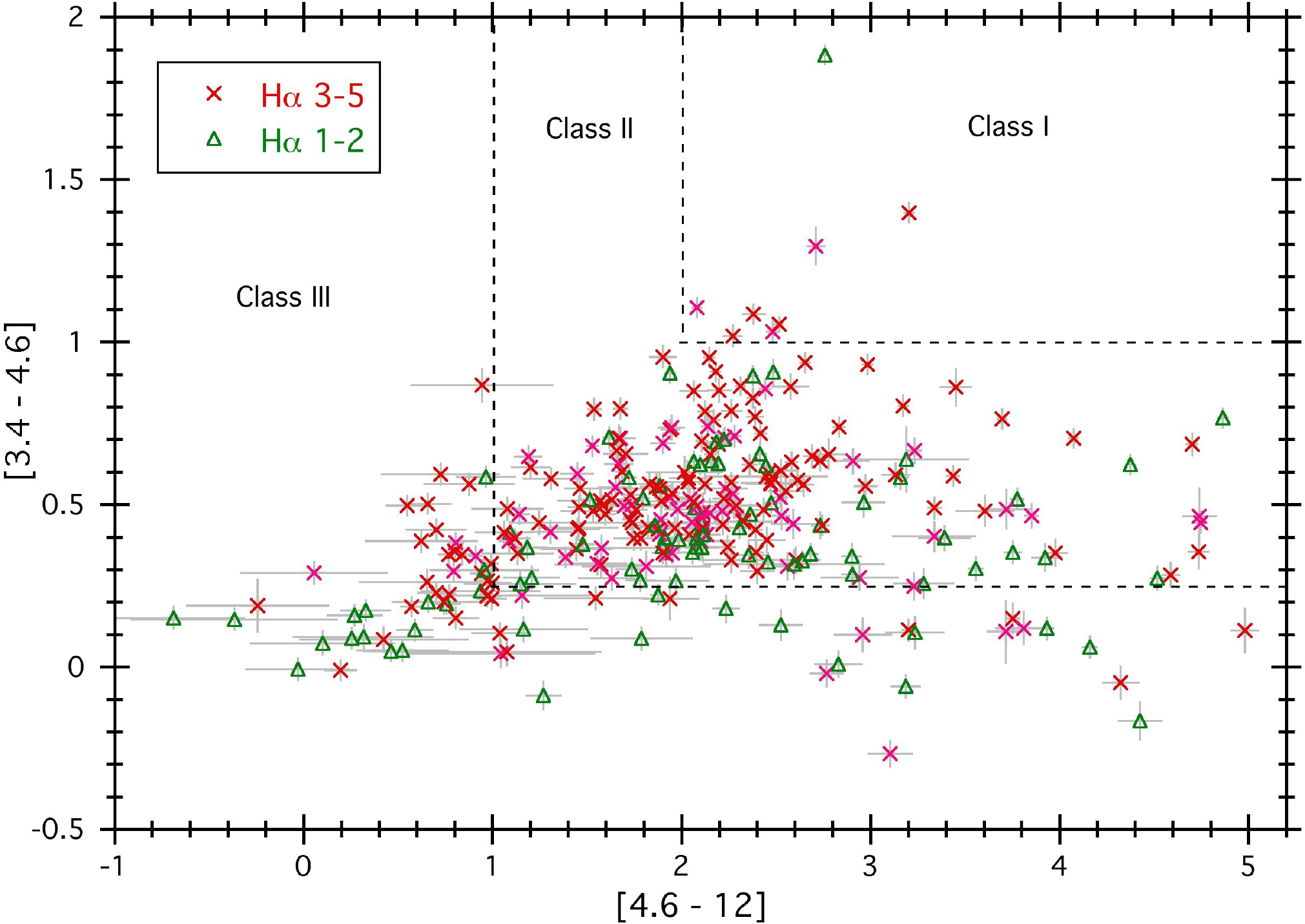}
      \caption{Distribution in a WISE two-colour diagram, of weaker-lined stars 
      (green triangles) and stronger-lined stars (red crosses).
        The dashed lines separate Class I, II and III sources, see  discussion in \citet{KoenLeis12}  }
         \label{fig8}
   \end{figure}

Among the 398 H$\alpha$ emitters listed here, 75\% were detected by
WISE (Wide-field Infrared Survey Explorer) with valid photometry in [3.4], [4.6] and [12] bands.  Figure~\ref{fig8} shows
a WISE [3.4-4.6] vs [4.6-12] colour-colour diagram. \citet{KoenLeis12}
 empirically determined the approximate boundaries of Class~I,
II, and III sources in this diagram. We have plotted our sources and
have marked strong H$\alpha$ emitters (group 3-4-5) as red crosses,
and weak H$\alpha$ emitters (group 1-2) as green triangles.  The 300
H$\alpha$ emitters that have WISE detections with valid photometry are
distributed across the diagram so 7 objects fall in the Class~I area,
221 fall in the Class~II area, and 72 fall in the Class~III area.  The
ratios of strong-to-weak H$\alpha$ emitters are 6:1 for Class~I
objects, 2.8:1 for Class~II objects, and 1.6:1 for Class~III
objects. The general expectation is that accretion declines with age
 \citep[e.g.][]{Bertout07}, and with it H$\alpha$ emission, so it is
gratifying to see that the ratio of strong-to-weak-line objects is
decreasing from Class~I to Class~III objects.

  \citet{FiscPadg16} have used WISE to survey an area of 100~deg$^2$
  around Canis Major for young stellar objects, an area almost ten
  times larger than our survey area. They find 479 YSOs, which they
  classify as 144 Class~I candidates and 335 Class~II candidates. Of
  their 479 objects, 295 fall within the area we have surveyed, but of
  these only about 1/3 (91 objects) are also detected in our survey.
  More recently, \citet{sewilo19} have used mid-infrared Spitzer and
  far-infrared Herschel data to identify further embedded sources in
  the Canis Major clouds, again with limited overlap with our survey.
  The explanation undoubtedly is that these infrared surveys by their
  nature are more sensitive to an embedded population, whereas our
  optical survey almost exclusively is focused on objects with low
  extinction. Thus the three surveys complement each other rather well.

\subsection{Gaia-DR2 distance}

There have been a number of attempts to determine the distance to the
CMa star-forming region. \citet{Racine68} used photometry and
spectroscopy of the stars with reflection nebulae identified by 
\citet{vdBergh66} to determine a distance of 690$\pm$120~pc.  \citet{Clar74y}
 studied the stars in the CMa~OB1 association and found a
distance of 1150$\pm$140~pc. \citet{ShevEzhk99} studied 88
likely OB stars in the CMa~OB association and found a distance of
1050$\pm$150~pc. Subsequently, \citet{KaltHild00} used
uvby$\beta$ photometry of the brighter OB stars to determine a closer
distance of 990$\pm$50~pc. Finally, \citet{LombAlve11} used the
observed density of foreground stars with predictions from a galactic
model to infer a distance of 1150$\pm$64~pc.

   \begin{figure}
   \centering
   \includegraphics[width=\columnwidth]{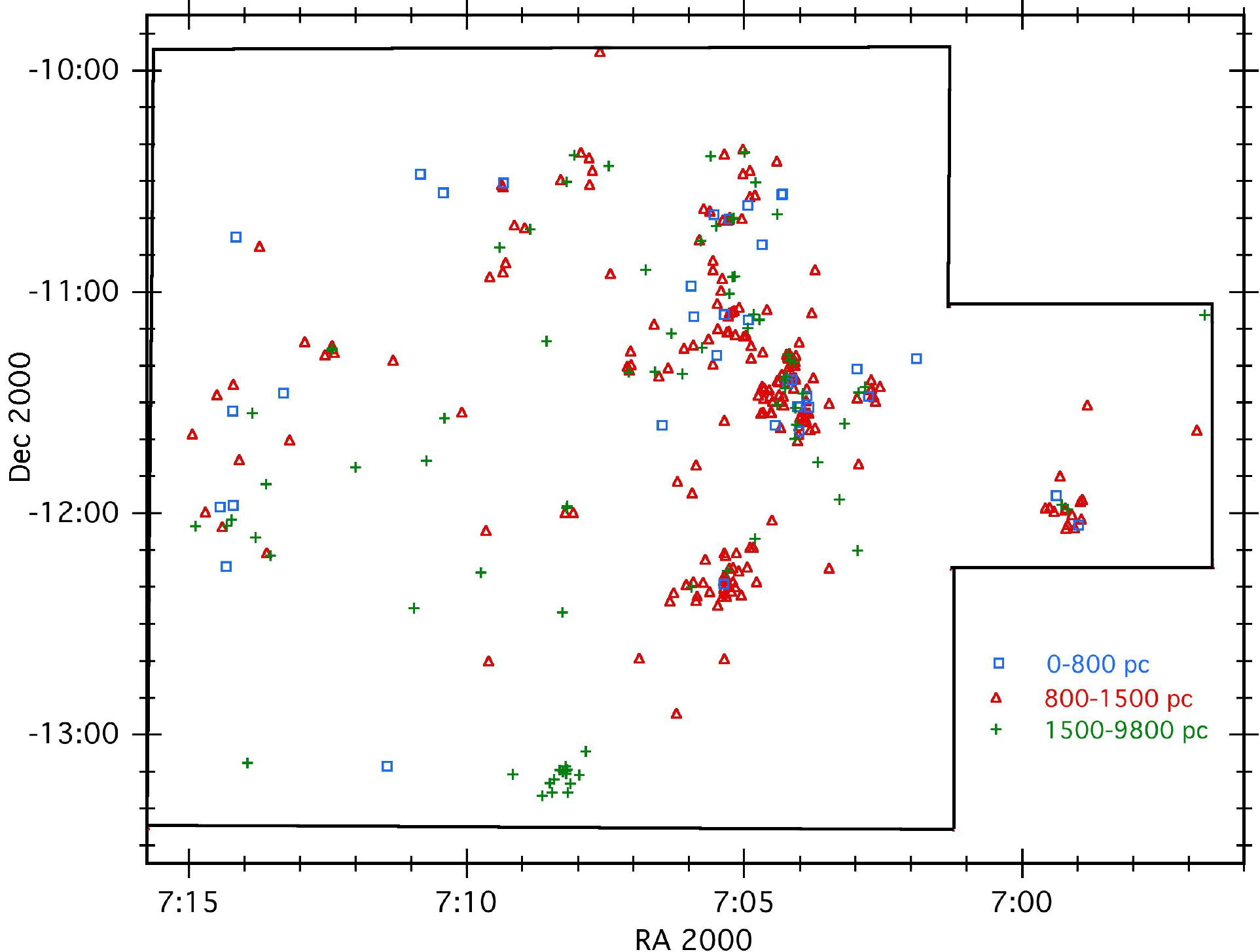}
      \caption{Plot of distribution of H$\alpha$ emission stars towards the Canis Major star-forming region. Only stars with positive parallaxes nearer than 10 kpc are shown, a  total of 327 stars. The stars are divided into nearby stars, from 0 to 800~pc, which are likely dMe dwarfs (blue), from 800 to 1500~pc, which include all the members of the Canis Major association (red), and stars more distant than 1500~pc, which are likely Be stars (green). The red points closely follow the gas surrounding the CMa shell. The small group of green symbols near the bottom of the figure marks the distant cluster NGC~2345.
              }
         \label{GaiaDR2_spatial}
   \end{figure}

\begin{figure}
   \centering
   \includegraphics[width=\columnwidth]{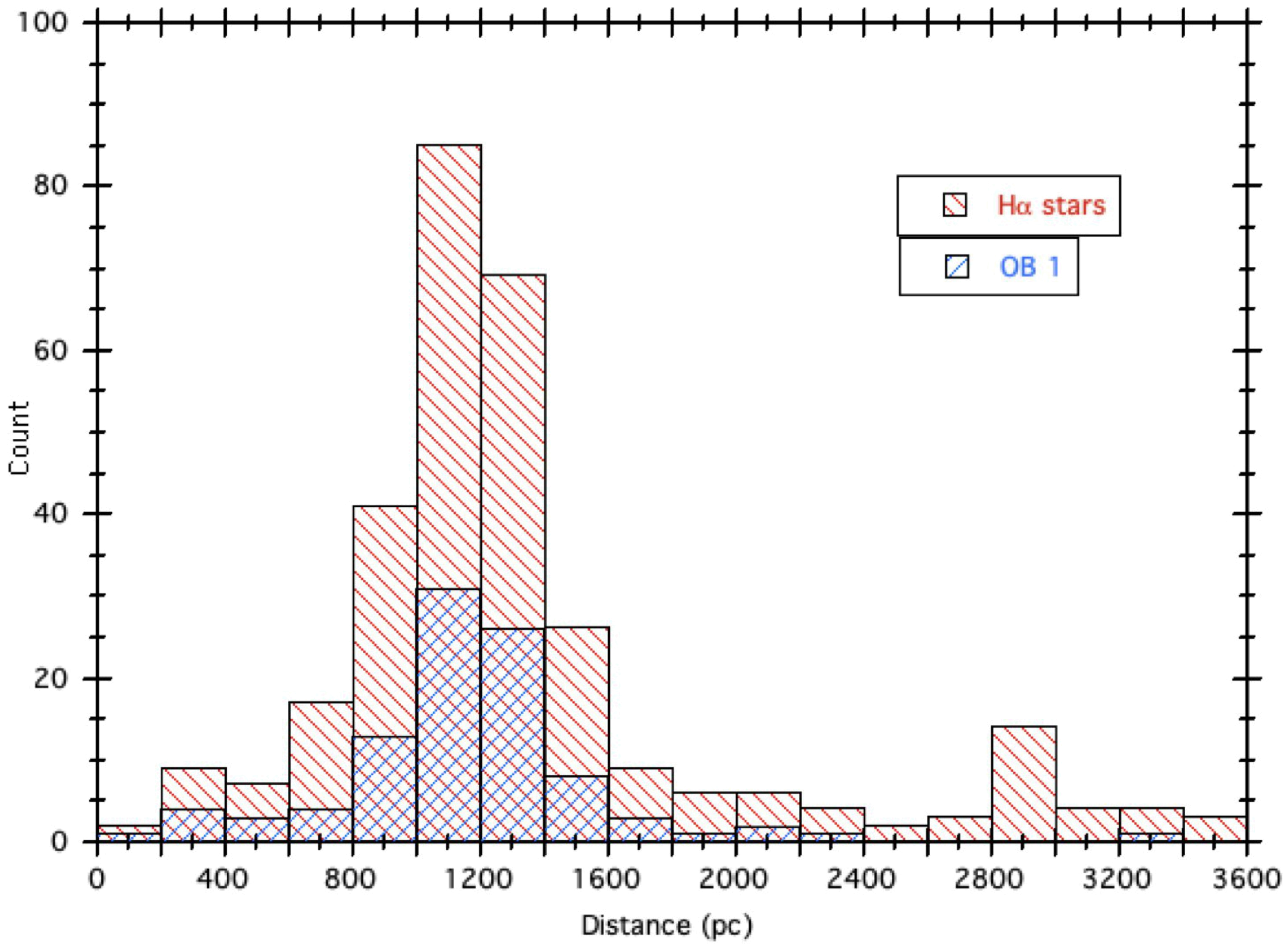}
      \caption{Histogram showing distribution of H$\alpha$ emitters (red) as function of distance with OB stars (blue) overlaid. The large peak corresponds to the Canis Major association, and the small peak at 2900~pc is the cluster NGC~2345.
              }
         \label{GaiaDR2_lt_3600_rev}
   \end{figure}

With our H$\alpha$ emission star survey, we can use the Gaia-DR2
catalogue to determine a more accurate distance to the CMa star-forming
region \citep{GaiaCol16,GaiaColl17}.
All but one of the 398 H$\alpha$ stars in Table~2 were found in the
Gaia-DR2 catalogue within a radius of one arcsec, but of these 28 have
no parallax, another 23 have negative parallaxes, and 13 had two
objects in the Gaia catalogue. All of these 64 cases were dismissed. The
projected spatial distribution of the remaining 334 stars, except 7 stars with distances larger than 10 kpc, is seen in
Fig.~\ref{GaiaDR2_spatial}, where stars in the interval from 0 to
800~pc are blue, from 800 to 1500 pc are red, and more distant stars
are green. The large majority of stars belong in the intermediate
interval, and show a distribution that closely follows the
distribution of gas in the region. The nearer and more distant objects
are likely dMe dwarfs and Be stars, respectively. The little clump of
stars at $\alpha$=7:08 and $\delta$=$-$13:10 is the more distant open cluster
NGC~2345. Figure~\ref{GaiaDR2_lt_3600_rev} shows a histogram of Gaia
distances for all H$\alpha$ emission-line stars (red). There is a
pronounced peak centred between
1.0 and 1.2 kpc corresponding to the CMa young stars, and a small peak
at $\sim$2900~pc from the $\sim$70~Myr old cluster NGC~2345, located
near the Perseus arm. 

In addition, in Fig.~\ref{GaiaDR2_lt_3600_rev} we have plotted the Gaia DR2 distances of all the OB stars (blue) that  \citet{Clar74y} and  \citet{ShevEzhk99} have suggested as members of the Canis Major OB association. It is seen that the distribution is almost identical to that of the H$\alpha$ emission-line stars. 

In order to refine the distance determination by eliminating the most
uncertain distances and very close and very distant stars, we have
selected a subset of H$\alpha$ emission-line stars that have parallax
errors less than 20\% and are located between 1050 and 1350~pc, a
total of 98 stars.  
The mean and median of this distribution
are 1184 and 1185~pc, respectively, with a standard deviation of 70~pc and
standard error of 7.1~pc. We have culled the
list of OB stars in a similar way to the H$\alpha$ emission-line stars.
The mean and median of the OB
distribution is 1204 and 1182~pc, respectively, with a 
standard deviation of 83~pc and a standard error
of 12~pc, that is, essentially the same as for the H$\alpha$
emission-line stars. These two independent measurements lend credence 
to the results.

The standard error is only meaningful if all stars were at the exact
same distance, but the CMa shell is about 50~pc in diameter, so
assuming an approximately spherical shape of the shell, a fundamental
uncertainty for the distance is $\sim$$\pm$25~pc, thus the best
distance estimate is 1185$\pm$25~pc.

\subsection{The runaway stars}

 \citet{HerbAsso77} suggested that the characteristic shell
structure of molecular clouds seen towards the CMa OB1 association is
the result of a supernova explosion. They identified a potential
runaway star, the O7V star HD~54662, which is the earliest and
brightest member of the CMa~OB1 association. However, subsequent
studies have cast doubt on its runaway nature, see \citet{MossMahy18} for a discussion. An alternative runaway candidate was
proposed by  \citet{ComeTorr98} who noted the peculiar motion of
the O9IV star HD~57682. We have extracted the Gaia-DR2 parallaxes for
these two stars to determine their distance and potential connection
to CMa~OB1. For HD~54662 we find a distance of 1170$^{+121}_{-100}$~pc, which
is identical within errors to the distance we find for the CMa
association, as one would expect for the brightest member of the
association. For HD~57682 we find a distance of 1241$^{+86}_{-76}$~pc, which
is consistent with the more uncertain photometric distance of
\citet{BergSchm96}. While this is somewhat further away than we
determined for the CMa association, both stars are within errors
compatible with our distance estimate. It follows that Gaia-DR2 does
not help to determine which of the two is the better runaway
candidate.  Finally, most recently \citet{fernandes19} have identified
a third runaway star in the region, HD 53974. Unfortunately, Gaia DR2
does not provide a parallax for this star. \citet{fernandes19} suggest
that the three runaway stars indicate that supernova explosions
occurred $\sim$6 Myr, $\sim$2 Myr and $\sim$1 Myr ago, leading to several 
populations of young stars.

\subsection{Herbig-Haro objects}

\begin{figure}
   \centering
   \includegraphics[width=74 mm]{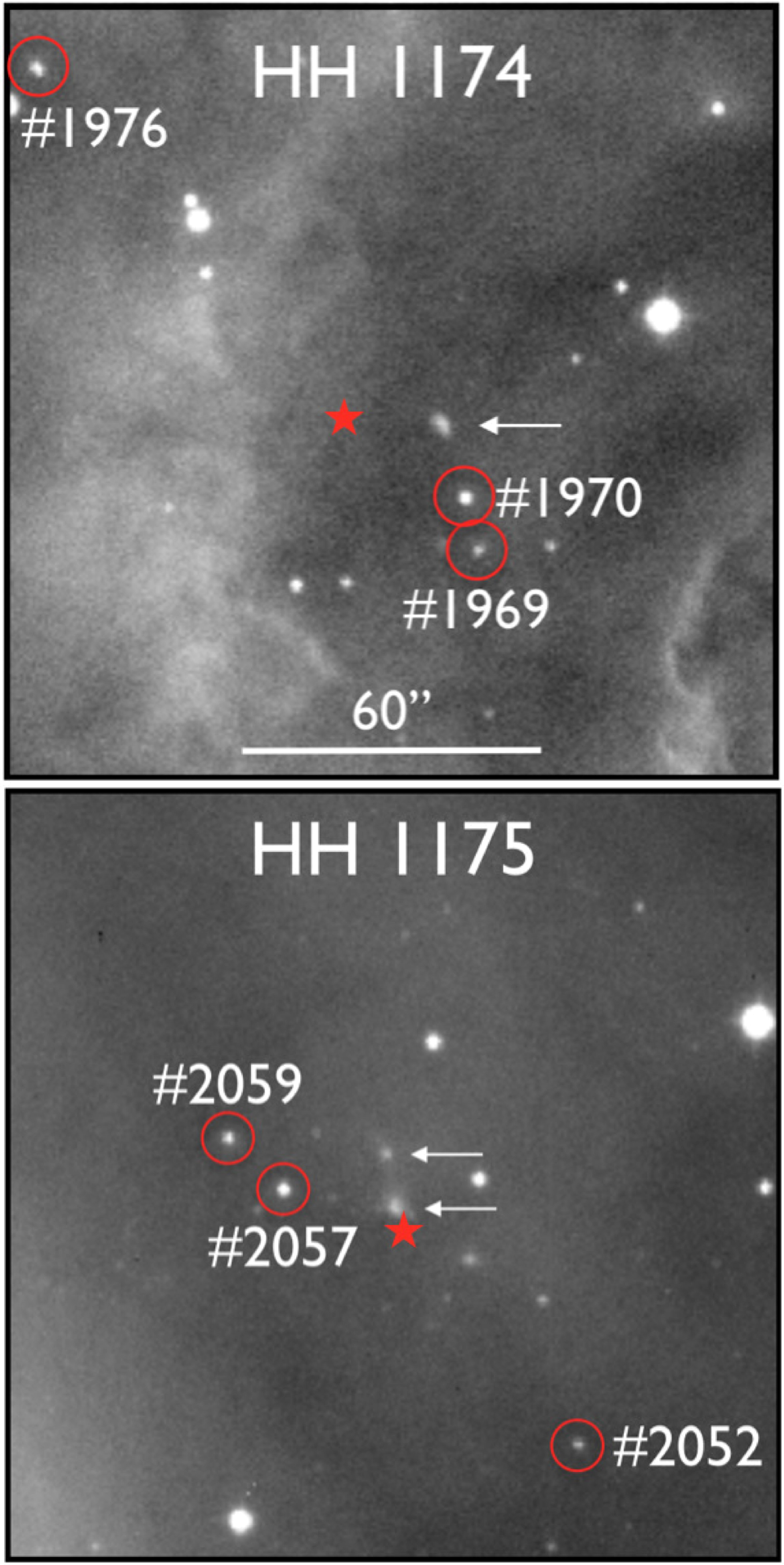}
      \caption{The two Herbig-Haro objects discovered in this
survey, HH~1174 and HH~1175, are seen in these H$\alpha$ images, each
150~arcsec wide, corresponding to $\sim$0.85~pc. The HH objects are
marked with arrows, and surrounding H$\alpha$ emission stars are
indicated by red circles. The probable sources of the two HH flows are
marked with red asterisks. Images courtesy John Bally.
}
         \label{hh}
   \end{figure}

In the course of our survey we noted two new Herbig-Haro objects with
their characteristic H$\alpha$ and [SII] lines. Both objects are
located towards high extinction regions and are surrounded by
H$\alpha$ emission-line stars. HH~1174 consists of a single knot,
located at 7:04:44.3 -11:07:16 (2000). This is 19~arcsec (0.1~pc) from
the IRAS source 07024-1102, which has colours similar to an
ultracompact HII region \citep{SlysDzur94}, and is associated with an
H$_2$O maser 
 \citep{BranCesa94} 
 and a dense $^{13}$CO core \citep{WangWu09}. 
The source is marked as a red asterisk in the upper panel
of Fig.~\ref{hh}, which is an H$\alpha$ image of the region. It
seems likely that the HH object is associated with this embedded
source.

HH~1175 consists of two knots, the coordinates of the brighter one is
7:05:35.4 -10:38:18 (2000). This is 3~arcsec from the embedded source
[EFM2013] G224.00452-01.73126 detected by \citet[their source
\#619]{EliaMoli13} using Herschel, and along the line defined by the
two knots, strongly suggesting that this is the driving source of the
HH flow. \citet{EliaMoli13} determine a bolometric luminosity of
6.4~L$_\odot$ (converted to our distance of 1185~pc), and classify it
as a protostellar object. The source is marked with a red asterisk in
the lower panel of Fig.~\ref{hh}.

\section{Conclusions}

We have performed a deep objective-prism survey for H$\alpha$ emission
line stars in a region of about 14 square-degrees covering mainly the
Canis Major OB1 region. We have obtained the following results.

1) A total of 398 H$\alpha$ emitters were detected, of which 353 are
newly discovered. A significant concentration is found associated with
the western side of the large shell of clouds surrounding the CMa~OB1
association, and they are likely to be mostly young stars. An
additional population of H$\alpha$ emitters is spread over the entire
region surveyed, and probably represents a mix of young stars with
foreground dMe stars and background Be stars. 

2) All but four of the stars are detected by 2MASS, and the
J-magnitude distribution shows a sharp cut-off around J~=~15, which is
the likely completeness limit. The H$\alpha$ emission-line strength
from 1 (weak) to 5 (strong) shows a broad peak around 3 and 4. 

3) When the H$\alpha$ emitters are divided into weak (1-2) and strong
(3-4-5) emitters and plotted against the visual extinction along each
sight line, no correlation is found such that strong H$\alpha$
emitters are more closely associated with higher extinction regions,
as one might otherwise have expected.

4) Placed in a J-H vs H-K$_s$ colour-colour diagram, about 44\% of the
H$\alpha$ emitters are found to have a near-infrared excess. This
compares to only 16\% determined in a similar survey of an area around
the very young Orion Nebula Cluster, possibly suggesting that the stronger UV
radiation near the ONC erodes the circumstellar matter more than in
the CMa~OB1 region. 

5) The subset of H$\alpha$ emitters with near-infrared excess are
found to be more likely to have strong rather than weak H$\alpha$
emission (ratio 3.4) than the entire population of H$\alpha$ emitters
(ratio 2.3).

6) 300 of the 398 H$\alpha$ emitters were detected by WISE, and when
plotted in a calibrated [3.4-4.6] vs [4.6-12] colour-colour diagram it
is found that 7 objects fall in the Class~I area, 221 fall in the
Class~II area, and 72 fall in the Class~III area. The ratios of
strong-to-weak H$\alpha$ emitters for Class~I, II, and III objects are
6, 2.8, and 1.6, respectively.

7) All but one of the 398 H$\alpha$ emitters were found in the
   Gaia-DR2 catalogue, and from these 327 with positive parallaxes were
   selected. Among these, 98 stars with parallax errors less than 20\%
   and with nominal distances from 1050 to 1350~pc were selected,
   showing a well-defined peak at a median distance of 1185~pc. The
   standard error on this is 7.1~pc, but since the extent of the
   association in the plane of the sky is about 50~pc, a more
   realistic value for the Canis Major association is 1185$\pm$25~pc.

8) We have used the compilations of OB stars towards Canis Major
   presented by \citet{Clar74y} and Shevchenko et al. (1999) and have
   selected 51 stars which have Gaia-DR2 parallax errors of less than 20\%
   and nominal distances in the range 1050 to 1350~pc, similar to the
   H$\alpha$ emission-line stars. The median distance for these OB
   stars is 1182$\pm$12~pc, in excellent correspondence with the value
   found from the H$\alpha$ emission-line stars. We also extract
   Gaia-DR2 distances to the two runaway candidate O-stars in the
   region, finding values of $1170^{+121}_{-100}$~pc for HD~54662, and
   $1241^{+86}_{-76}$~pc for HD~57682, confirming that both stars are
   members of the CMa~OB1 association.

9) While searching for the H$\alpha$ emission stars, we discovered two
   Herbig-Haro objects, here named HH~1174 and 1175, located towards high
   extinction regions of the L1657 cloud, and we identified embedded
   sources that are likely their driving sources.

\begin{acknowledgements}

We are grateful to Guido and Oscar Pizarro, who as telescope operators
obtained all the films used in this survey, and to John Bally for
obtaining the images in Fig.~\ref{hh}.  We acknowledge use of the
Digitised Sky Survey, which was produced at the Space Telescope
Science Institute under U.S. Government grant NAG W-2166.  
This work has made use of data from the European Space Agency (ESA) mission
{\it Gaia} (\url{https://www.cosmos.esa.int/gaia}), processed by the {\it Gaia}
Data Processing and Analysis Consortium (DPAC,
\url{https://www.cosmos.esa.int/web/gaia/dpac/consortium}). Funding for the DPAC
has been provided by national institutions, in particular the institutions
participating in the {\it Gaia} Multilateral Agreement.
This publication makes use of data products from the Wide-field Infrared
Survey Explorer and from the Two Micron All Sky Survey.  This research
has made use of the SIMBAD database, operated at CDS, Strasbourg,
France, and of NASA's Astrophysics Data System Bibliographic Services.

\end{acknowledgements}

{\em Facility:} ESO 1m Schmidt telescope, Gaia, 2MASS, WISE

\vspace{0.3cm}

\bibliographystyle{aa}
\bibliography{CMa}

\vspace{0.3cm} 

\appendix

Appendix A: Finding Charts

In this appendix we provide finding charts for all of the new H$\alpha$
emission-line stars presented in this paper. The image stamps, each 90$"$
$\times$ 90$"$, were obtained from red DSS2 images.

   \begin{figure*}
   \centering
   \includegraphics[width=15.5cm]{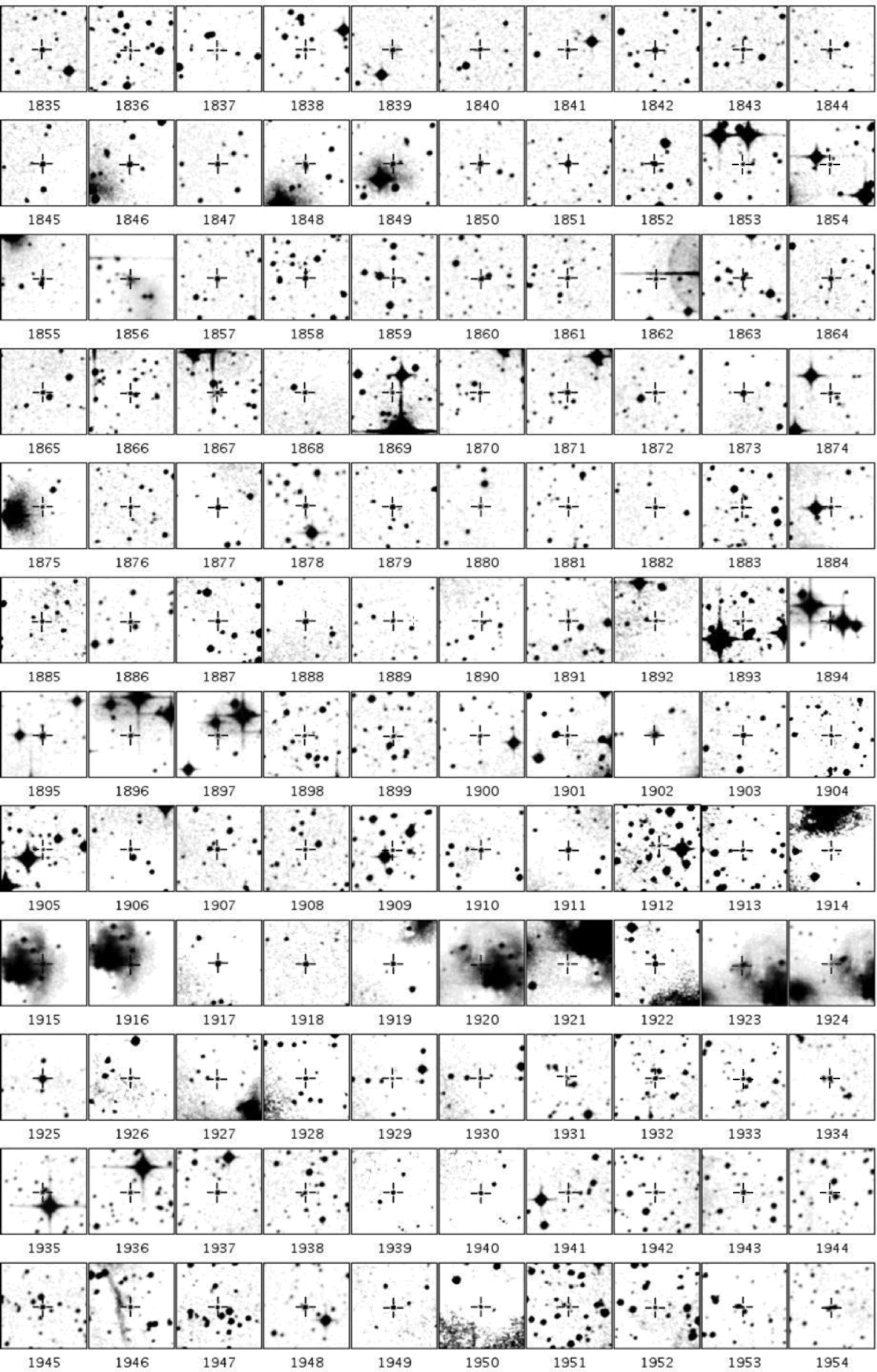}
      \caption{Finding charts, 90\"\ to a side.}
         \label{figcharts}
   \end{figure*}

   \begin{figure*}
   \centering
   \includegraphics[width=15.5cm]{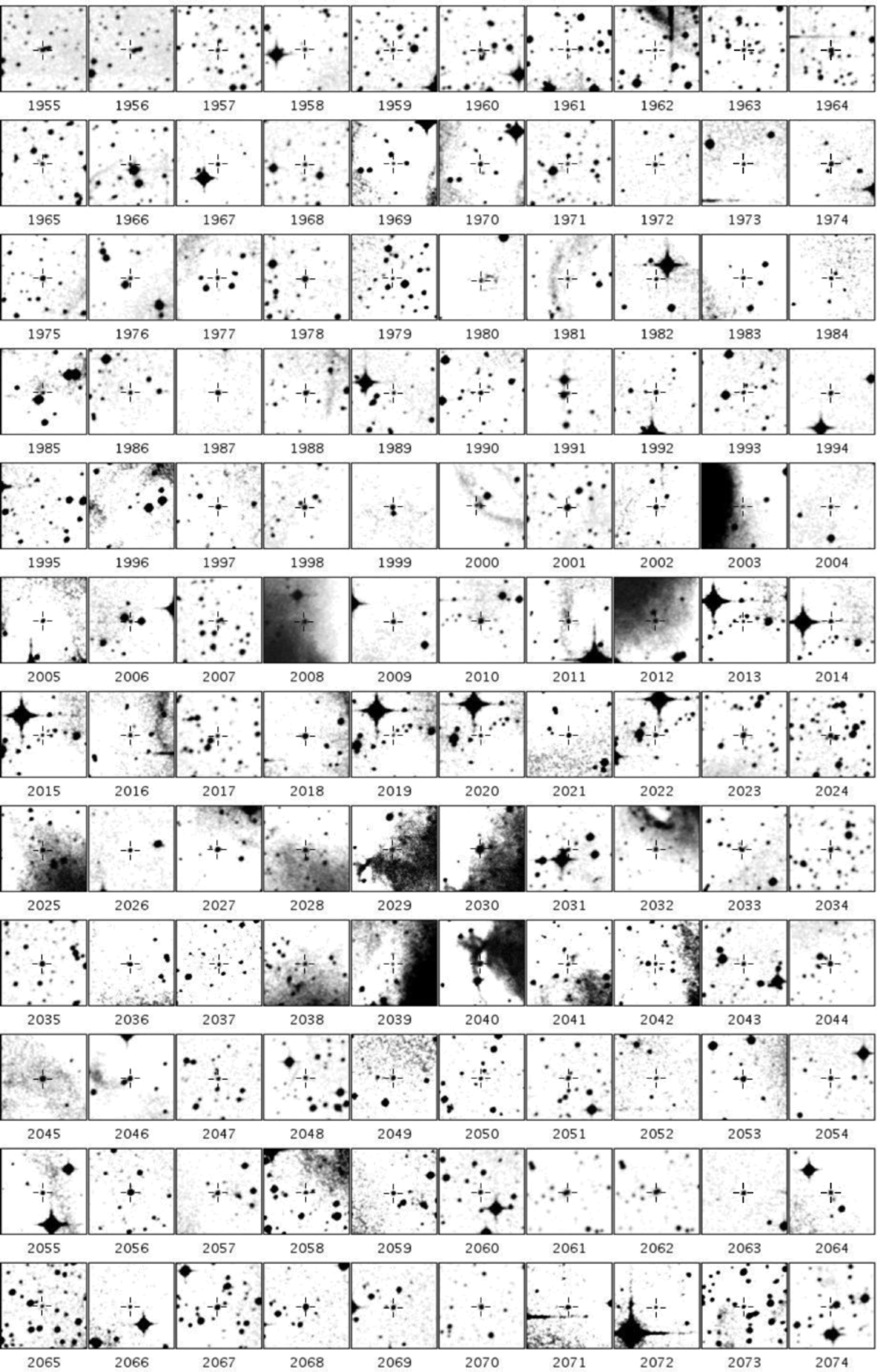}
      \caption{Finding charts, 90\"\ to a side.}
   \end{figure*}

   \begin{figure*}
   \centering
   \includegraphics[width=15.5cm]{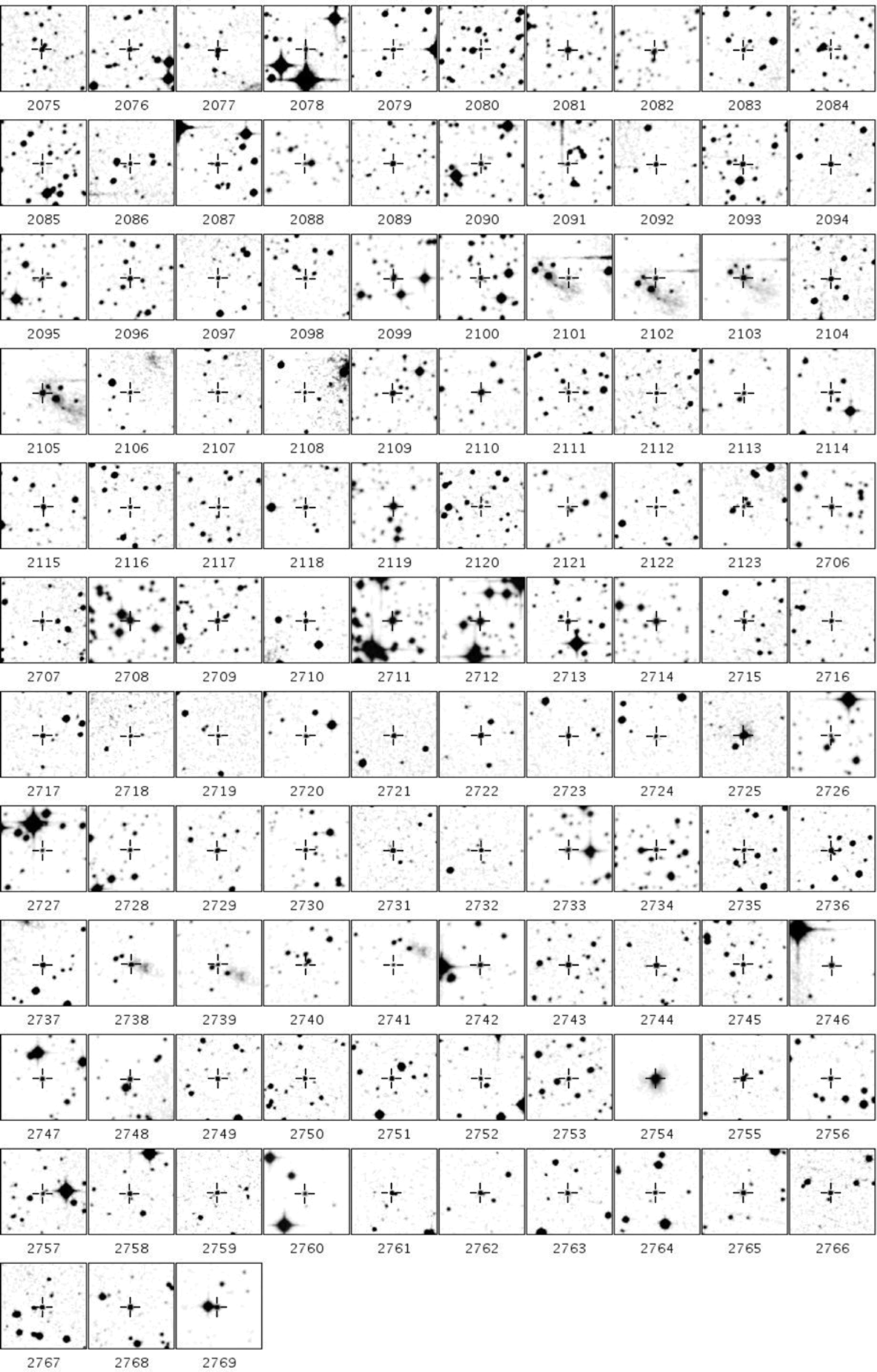}
      \caption{Finding charts, 90\"\ to a side.}
   \end{figure*}

\end{document}